\documentclass{nature}
\usepackage{latexsym}
\usepackage{color}
\usepackage{graphicx}
\usepackage{fancyheadings}
\newcommand\nodata{ ~$\cdots$~ }%
\usepackage{lscape}

\title{Jet suppression by accretion disk winds in the microquasar GRS 1915+105}

\author{Joseph Neilsen$^{1}$ \& Julia C. Lee$^{1}$}

\begin{document}

\maketitle

\begin{affiliations}
 \item Harvard University Department of Astronomy, 60 Garden Street, MS-10, Cambridge, MA 02138
\end{affiliations}

\begin{abstract}
Stellar-mass black holes with relativistic jets, also known as
microquasars, mimic the behaviour of quasars and active galactic
nuclei$^{1}$. Because timescales around stellar-mass black holes are 
orders of magnitude smaller than those around more distant
supermassive black holes, microquasars are ideal nearby `laboratories'
for studying the evolution of accretion disks and jet formation in
black-hole systems$^{2}$. Although studies of black holes have revealed a
complex array of accretion activity, the mechanisms that trigger and
suppress jet formation remain a mystery. Here we report the discovery
of a broad emission line during periods of intense hard X-ray flux in the
microquasar GRS 1915+105, and highly ionized narrow absorption lines
during softer states. We argue that the broad emission line
arises when the inner accretion disk is illuminated by hard X-rays,
possibly from the jet$^{3}$. In contrast, during softer states, when
the jet is weak or absent$^{4}$, absorption lines appear
as the powerful radiation field around the black hole drives a hot
wind off the accretion disk$^{5-7}$. Our analysis strongly suggests
that this wind carries enough mass away from the disk to halt the flow
of matter into the radio jet.
\end{abstract}

GRS 1915+105 is a 14 $M_{\odot}$ black hole accreting matter from a
0.8 $M_{\odot}$ K3 IV star in a wide 33.5-day orbit$^{8}$ (here
$M_{\odot}$ is the solar mass). As the
first known source of superluminal jets in the Galaxy$^{9}$, with a
lightcurve exhibiting at least 14 distinct classes of
high-amplitude variability due to rapid disk-jet interactions$^{3,10-15}$, this
microquasar provides a fascinating example of the coupling between
jets and accretion disks around black holes. To study this
relationship, we have analysed archival HETGS$^{16}$ (High Energy
Transmission Grating Spectrometer) observations of GRS 1915+105 from
the Chandra X-ray Observatory. Between 2000 April 24 and 2007 August
14, the HETGS observed this mircroquasar 11 times with high spectral
resolution, probing five of the fourteen variability classes of this
enigmatic X-ray source. The data include six observations of the
faint, hard, jet-producing state$^{17}$ (Obs.\ H1--H6) and five
observations of various bright, softer states (Obs.\ S1--S5).

As one of the brightest X-ray sources in the sky, GRS 1915+105
requires Chandra observations in a special high time-resolution mode
to mitigate photon pileup (that is, two or more photons striking a single 
pixel during one CCD frame time). Unfortunately, this mode is not
at present well calibrated, so it is currently impossible to fit a
`physical' model (e.g.\ a disk blackbody) to the broadband
continuum. However, because we are mainly interested in spectral
lines, we circumvent the calibration problems by fitting the X-ray
continuum with a smooth 10th-order polynomial, revealing spectral
features with widths less than 0.8 \AA. The fraction of the 0.7--4.1
\AA~(3--18 keV) flux emitted below 1.4 \AA~(above 8.6 keV), which we
call the hard flux fraction $HF$, clearly delineates the hard and soft
states (see Fig.\ 1) and the observed spectral features (Fig.\ 2 and
Supplementary Information). We will argue that these spectral
differences are strong X-ray indicators of the disk-jet coupling in
GRS 1915+105.

The states with high $HF$ are dominated by a broad emission line near
1.86 \AA~(6.7 keV), which we tentatively identify as Fe\,{\sc xxv}
(Table 1). In contrast, strong narrow absorption lines near 1.77
\AA~(7.0 keV), consistent with Fe\,{\sc xxvi} absorption blueshifted
by $\sim1,000$ km s$^{-1}$, are seen in softer states. In some cases,
a weaker Fe\,{\sc xxv} absorption line is seen at the same
velocity. The absence of charge states other than hydrogen-like and
helium-like iron in Obs.\ S2--S5 indicates a highly ionized absorber
with ionization parameter $\xi=L_{\rm X}/nr^{2}$ of the order $10^{4}$
(here $L_{\rm X}$ is the X-ray luminosity, $n$ is the gas density, $r$
is the distance from the source of ionizing radiation, and
$\xi=10^{4}$ corresponds to $T\sim10^{6}$ K)$^{18}.$ These results are
consistent with previous X-ray spectral studies of GRS
1915+105$^{19-21}$.

It has been postulated that the accretion disk is truncated at some
distance from the black hole during the low hard state$^{3,22}$, when
the spectrum may be dominated by hard X-rays from the corona or
jet$^{4,23}$. Coupled with infrared studies implicating jet activity in the
production of emission lines in GRS 1915+105$^{24}$, we conclude that
the broad Fe\,{\sc xxv} emission line is produced when the inner edge
of the disk is illuminated by these hard X-rays. Our interpretation is 
substantiated by the fact that the equivalent width of the 
Fe\,{\sc xxv} emission line increases with both $L_{\rm X}$ and the
radio flux at 15 GHz as measured by the Ryle  Telescope$^{15}$ (Fig.\
3). Under the assumption that the line broadening $(\geq12,500$ km s$^{-1}$)
is due to orbital motion in a Keplerian disk, the line is emitted at
$r\leq1.1\times10^{9}$ cm $(\leq255~R_{\rm S},$ where $R_{\rm S}$ is the
Schwarzschild radius of the black hole). This is a reasonable upper
limit for the inner edge of the truncated disk.
 
In comparison, the inner edge of the accretion disk may lie much
closer to the black hole (as little as $r=3R_{\rm S}$ or less) during
bright soft states$^{25}$. In this context, we suggest that the
absorption lines seen in GRS 1915+105 originate in an accretion
disk wind. Several lines of evidence support this interpretation: the
absorption lines (1) only appear during softer states, when the disk
may be prominent and hard X-ray illumination is relatively weak (Fig.\
4), (2) are all narrow and blueshifted (see Table 1), implying
material flowing into our line of sight, and are accompanied by (3) a
weak emission line at a slightly longer wavelength (Fig.\
2). Together, (2) and (3) constitute a P-Cygni profile, a classic wind
signature. Since a K-type companion cannot drive a strong wind, the
wind must originate in the accretion disk. Furthermore, the wind speed
corresponds to the escape velocity from the black hole at a distance
of $r=2.5\times10^{11}$ cm ($=53460~R_{\rm S}$), which is well inside
the accretion disk of GRS 1915+105. 

Given the high luminosity of this black hole binary and the strong
variation of the equivalent width of the Fe\,{\sc xxvi} absorption
line with $L_{\rm X}$ (Table 1), it is likely that radiation pressure
plays a role in driving this highly ionized wind$^{7}.$ However,
radiation pressure alone (mainly imparted by ultraviolet 
emission lines) is inefficient at transferring momentum to a wind at
$\xi>10^{3}$ (ref 26). But at the high luminosity of GRS 1915+105,
X-ray heating and thermal pressure can provide the extra boost to
drive a hot, fast wind off the accretion 
disk for $r>0.01r_{\rm C},$ where the critical radius $r_{\rm C}$ is
given by $r_{\rm  C}= (9.8\times10^{9})\times (M_{\rm BH}/M_{\odot})
/T_{\rm C8}$ cm. Here $M_{\rm BH}$ is the black-hole 
mass, and $T_{\rm C8}$ is the electron temperature in units of
$10^{8}$ K (refs 5,6). As our analysis indicates that the wind
temperature $T_{\rm wind} \sim10^{6}$ K, the wind could originate at
any $r>1.4\times10^{11}$ cm ($=29340~R_{\rm S})$. Although the
relevant electron temperature could be much higher$^{5},$ this
launching radius is consistent with our earlier estimate from the
blueshift of the wind, so that thermal driving assisted by radiation
pressure successfully explains the origin of this wind.

To estimate the mass loss rate from the wind, we fit the
spectra with a photoionized absorption model that calculates the
ionization balance for a shell of gas surrounding a central
X-ray source$^{18}$. This model is characterized mainly by the
absorbing gas column density and ionization parameter $\xi\sim10^{4};$
for a fixed luminosity and wind speed, a higher ionization
parameter implies a smaller mass loss rate$^{19}$. Assuming an accretion
efficiency $\eta,$ accretion rate $\dot M_{\rm acc}$, and wind
covering factor $f,$ our model implies a wind mass loss rate of $\dot
M_{\rm W}=188\dot  M_{\rm acc}~\eta~f.$ With $f<5\%$ (Fig.\ 2 legend)
and $\eta=6\%$ (ref 27), we calculate $\dot M_{\rm W}<0.59~\dot M_{\rm
  acc}$ ($\sim10^{-8}$ M$_{\odot}$ yr$^{-1}$). 

Interestingly, this wind drives approximately the same mass loss as
the radio jet$^{17},$ suggesting that GRS 1915+105 is able to maintain 
a rough equilibrium between mass accretion and outflow, independent of
its spectral state and the outflow mechanism, over the span of our 
observations. Furthermore, the noticeable decrease of $HF$ with the
equivalent width of the absorption lines (Fig.\ 4) indicates a complex
competition between the accretion disk wind and the radio jet. When $HF$
decreases, there are fewer hard X-rays available to over-ionize the
wind, allowing it to carry away more of the matter that sustains the
jet. Thus it appears that Comptonization and photoionization mediate
the coupling between the jet and the disk in GRS 1915+105.

This is a strong indication that like their supermassive counterparts,
stellar-mass black holes can regulate their accretion rate by feedback
into their environments. More importantly, these observations clearly
demonstrate that at sufficiently high luminosities in GRS 1915+105,
the intense radiation field of the disk redirects the accretion flow,
away from the radio jet and into a wind. By revealing a surprisingly
simple jet-quenching mechanism in GRS 1915+105, our results point to
fundamental new insights into the long-term disk-jet coupling around
accreting black holes and hint at tantalizing evidence of the
mechanism by which stellar-mass black holes regulate their own
growth.

\begin{addendum}
\item[Supplementary Information] is linked to the online version of
  the paper at www.nature.com/nature. 
 \item We gratefully acknowledge support from the Harvard University
   Graduate School of Arts and Sciences (J. N.) and Faculty of Arts
   and Sciences (J.C.L.). We thank G. Pooley for providing the radio
   data used in this paper and we acknowledge conversations with
   R. Remillard, who provided the Rossi X-ray Timing Explorer spectra,
   and M. Begelman.
 \item[Author Contributions] J.N. processed the data, performed
   spectral analysis, and wrote the paper. J.C.L. commented
   extensively on the manuscript. Both authors discussed the results
   at length.
 \item[Author Information] Reprints and permissions information is
   available at npg.nature.com/reprintsandpermissions. The authors
   declare that they have no competing financial
   interests. Correspondence and requests for materials should be
   addressed to J.N. (jneilsen@cfa.harvard.edu).
\end{addendum}
\begin{table}
\caption{Spectral properties of GRS 1915+105}
\scriptsize
\begin{tabular}{lccccccccccc}
\hline
\hline
Obs.\ & Obs.\ & X-Ray & &  & $S_{\rm 15~GHz}$ & Line &
$\lambda_{0}$ & $\lambda_{\rm obs}$ & $\Delta v$ & $W_{0}$ & $\sigma$ \\
\# & ID & State$^{10}$ & $L_{38}$ & HF & (mJy) & ID &
(\AA) & (\AA) & (km s$^{-1}$) & (eV) & (km s$^{-1}$)\\
\hline
S1 & 7485 & $\phi$ &  3.1 & 0.14 & \nodata & Fe\,{\sc xxvi} & 1.7807 & $1.7774_{-0.0005}^{+0.0004}$ & $ -560_{-80}^{+70}$ & $-29.9_{-1.6}^{+1.3}$ & $ 650_{- 160}^{+150}$ \\
S2 & 6581 & $\gamma$ & 12.8 & 0.18 & $5\pm3$ & Fe\,{\sc xxvi} & 1.7807 & $1.775\pm0.001$ & $-1000_{-220}^{+240}$ & $-21.9_{-2.7}^{+2.2}$ & $1160_{-250}^{+280}$ \\
S3 & 1945 & $\rho$ &  4.9 & 0.21 & $3\pm2$ & Fe\,{\sc xxvi} & 1.7807 & $1.772\pm0.002$ & $-1420_{-310}^{+320}$ & $-7.2\pm1.7$ & $<1120$ \\
S4 & 6580 & $\beta$ &  5.9 & 0.21 & 20--60 & Fe\,{\sc xxvi} & 1.7807 & $1.774\pm0.002$ & $-1100_{-300}^{+360}$ & $-19.3_{-3.5}^{+3.2}$ & $980_{-420}^{+450}$ \\
S5 & 6579 & $\beta$ &  5.2 & 0.21 & 20--60 & Fe\,{\sc xxvi} & 1.7807 & $1.775_{-0.003}^{+0.002}$ & $ -910_{-430}^{+390}$ & $-13.3_{-2.9}^{+3.0}$ & $<1080$ \\
H1 & 660 & $\chi$ &  3.0 & 0.31 & $20\pm4^{19}$ & Fe\,{\sc xxvi} & 1.7807 & $1.775_{-0.003}^{+0.004}$ & $ -910_{-570}^{+630}$ & $-4.8_{-2.5}^{+1.6}$ & $<1300$ \\
H1 & 660 & $\chi$ &  3.0 & 0.31 & $20\pm4^{19}$ & Fe\,{\sc xxv} &  1.868 & $ 1.89\pm0.02$ & $ 2830_{-3840}^{+3830}$ & $53.0\pm7.9$ & $17820_{-2950}^{+2000}$ \\
H2 & 4587 & $\chi$ &  3.9 & 0.32 & $89\pm8$ & Fe\,{\sc xxv} &  1.868 & $ 1.865\pm0.006$ & $ -460\pm980$ & $101.1\pm6.4$ & $11120_{- 750}^{+ 840}$ \\
H3 & 1944 & $\chi$ &  3.1 & 0.32 & $29\pm1$ & Fe\,{\sc xxv} &  1.868 & $ 1.867\pm0.009$ & $ -260_{-1480}^{+1460}$ & $81.9\pm7.1$ & $12210_{-1100}^{+1290}$ \\
H4 & 4589 & $\chi$ &  3.6 & 0.33 & $80\pm1$ & Fe\,{\sc xxv} &  1.868 & $ 1.861\pm0.007$ & $-1230_{-1150}^{+1180}$ & $91.8\pm6.9$ & $11130_{- 960}^{+1110}$ \\
H5 & 1946 & $\chi$ &  3.0 & 0.34 & $10\pm3$ & Fe\,{\sc xxv} &  1.868 & $ 1.86_{-0.02}^{+0.01}$ & $-2000_{-2340}^{+2300}$ & $53.6\pm7.4$ & $11950_{-1580}^{+1930}$ \\
H6 & 4588 & $\chi$ &  3.2 & 0.36 & $90\pm3$ & Fe\,{\sc xxv} &  1.868 & $ 1.867\pm0.008$ & $ -130_{-1240}^{+1230}$ & $82.1\pm6.8$ & $10590_{- 950}^{+1080}$ \\
\hline
\normalsize
\end{tabular}
The 11 HETGS observations are listed in order of increasing hard flux
fraction with their \textit{Chandra} observation ID 
numbers and relevant spectral properties. `S' indicates a soft state
and `H' indicates a hard state. The Greek letters identify
the observations with one of the 14 variability classes of this
microquasar$^{10}.$ $L_{38}$ is the 0.7-4.1
\AA~(3-18 keV) luminosity, measured with the Rossi X-ray Timing
Explorer (RXTE), in units of $10^{38}$ ergs s$^{-1}.$ HF is 
the hard flux fraction, defined as the ratio of the unabsorbed 0.7-1.4
\AA~(8.6--18 keV) to 0.7--4.1 \AA~(3--18 keV) continuum flux. $S_{\rm
  15~GHz}$ is the radio flux at 15 
GHz, measured by the Ryle Telescope$^{15}$. The line ID is our
identification of the strongest line; we detect Fe\,{\sc xxvi} in
absorption and Fe\,{\sc xxv} in emission. $\lambda_{0}$ is the ion's
laboratory wavelength and $\lambda_{\rm obs}$ is the observed
wavelength. $\Delta v$ is the corresponding Doppler velocity. $W_{0}$
is the line equivalent width, and $\sigma$ is the line width. All
errors correspond to 90\% confidence limits. Because of its
intermediate hard flux fraction, Obs.\ H1 exhibits both a weak
broad emission line and a weak narrow absorption line.
\label{table1}
\end{table}
\clearpage
\begin{figure}
\centerline{\includegraphics[width=\textwidth]{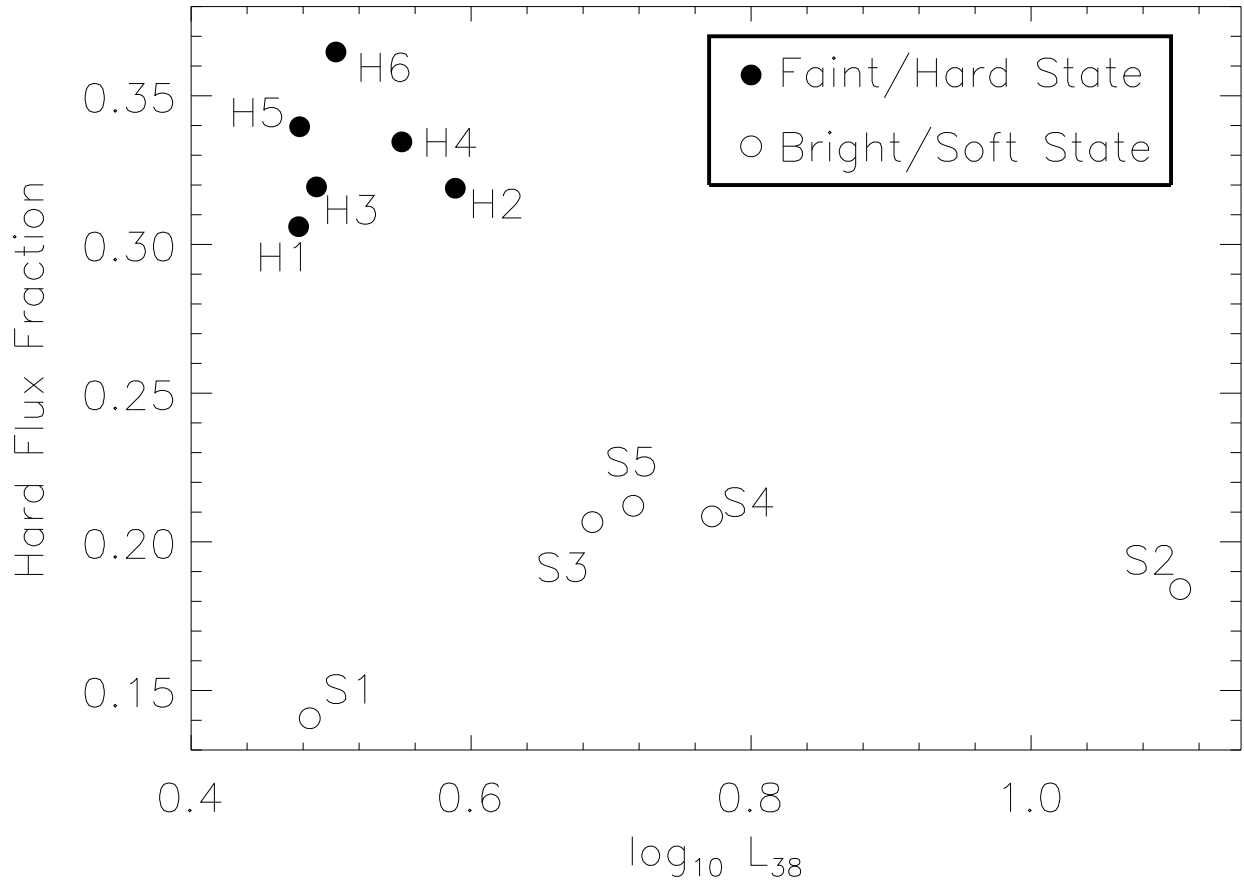}}
\caption{The X-ray luminosity and hard flux fraction for the 11
  archival HETGS observations of GRS 1915+105. $L_{38}$ is the X-ray
  luminosity in units of $10^{38}$ ergs s$^{-1},$ measured with RXTE
  from 0.7--4.1 \AA~(3--18 keV), assuming a distance of 12.5 kpc (ref
  9) and neutral hydrogen absorption $(N_{\rm H} =5\times10^{22}$ cm$^{-2})^{19}$
  along the line of sight. The hard flux fraction, used as a proxy
  for the strength of the Comptonized emission from the corona or jet,
  is defined as the ratio of the unabsorbed continuum flux from
  0.7--1.4 \AA~(8.6--18 keV) to 0.7--4.1 \AA~(3--18 keV). The 11
  observations are classified as hard or soft based on previous X-ray
  studies$^{10};$ as expected, the hard states have a higher hard flux
  fraction.}
\label{fig1}
\end{figure}
\begin{figure}
\centerline{\includegraphics[width=\textwidth]{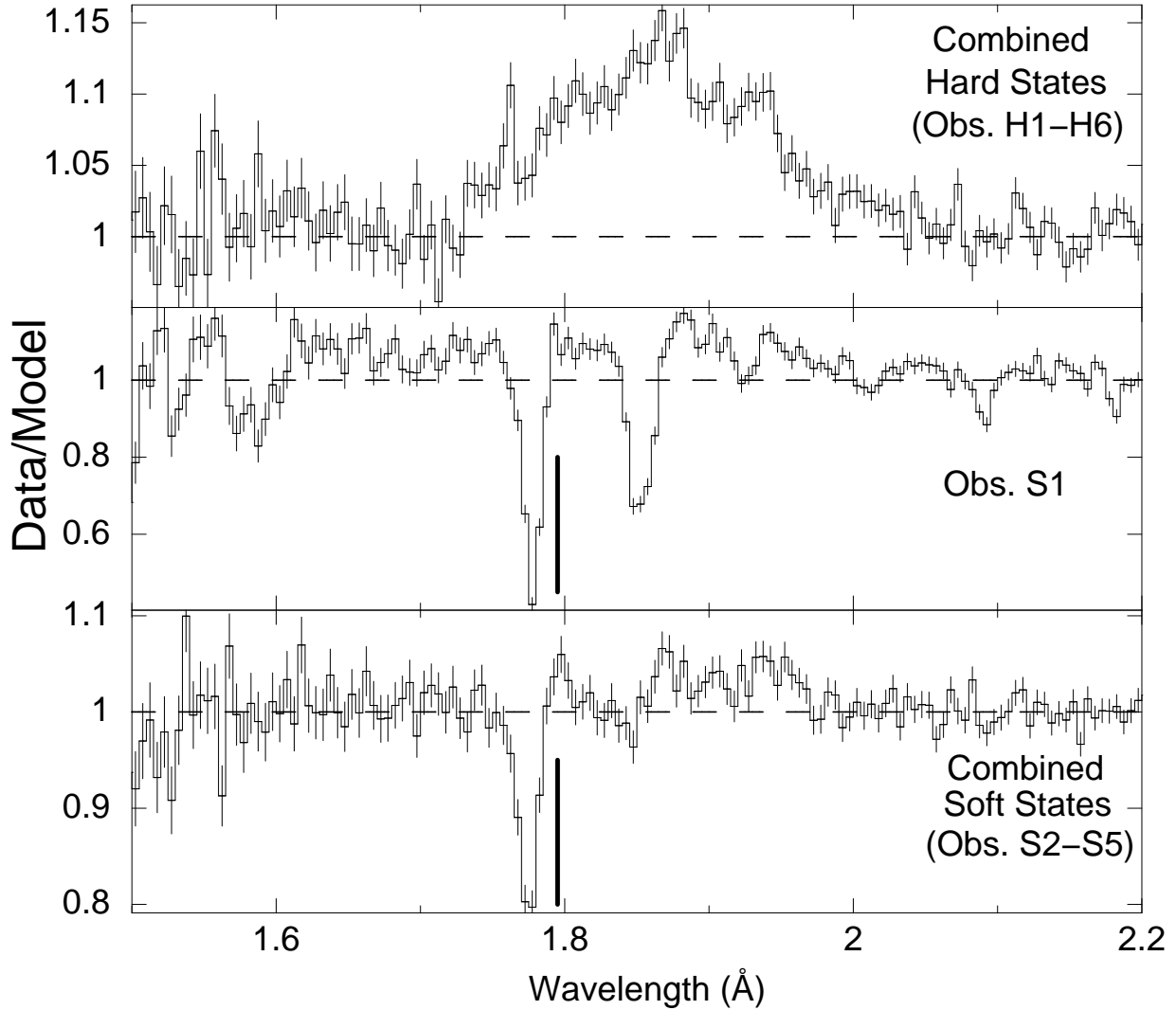}}
\vspace{-25mm}
\caption{The data:model ratio for the continuum fits to the HETGS
  observations of GRS 1915+105. We plot the combined hard states
  (Obs.\ H1--H6) and combined soft states (S2--S5) for clarity; Obs.\ S1 is
  shown alone to highlight other strong lines. Plots of the individual
  spectra can be found in the Supplementary Information. The broad
  Fe\,{\sc xxv} \textit{emission}
  line distinguishes periods of significant hard X-ray illumination
  from softer states, which are dominated by strong Fe\,{\sc xxvi}
  \textit{absorption} lines. We use a simple Gaussian to measure the properties
  of the broad Fe\,{\sc xxv} line (with the assumption that it is not
  a blend of emission from several Fe charge states) and find it has a
  line width $\geq 12,500$ km s$^{-1},$ much larger than
  the orbital velocity of either the companion star or the black
  hole$^{8}.$ We note that it represents a lower limit because our
  continuum-fitting procedure would mask the red wing of any
  relativistically broadened emission line; the implied inner disk
  radius of $255~R_{\rm S}$ is thus an upper limit. While the emission
  line width probes the size of the disk, the P-Cygni profile of the
  absorption lines constrains the geometry of the accretion disk
  wind. The vertical line indicates the position of the weak
  P-Cygni emission component. Because this emission component is weak,
  the wind must be confined to the equatorial plane of the disk. As
  the binary is viewed at an inclination $i=70^{\circ}$ (ref
  9), we can suppose that $i<20^{\circ}$ above the midplane of the disk is
  a reasonable estimate (implying $f<5\%$). Spectral analysis was
  performed with the ISIS$^{28}$ spectral fitting package. The errors
  shown are $68\%$ confidence limits on the data:model ratio.}
 \label{fig2}
\end{figure}
\vspace{15mm}
\begin{figure}
\centerline{\includegraphics[width=0.9\textwidth]{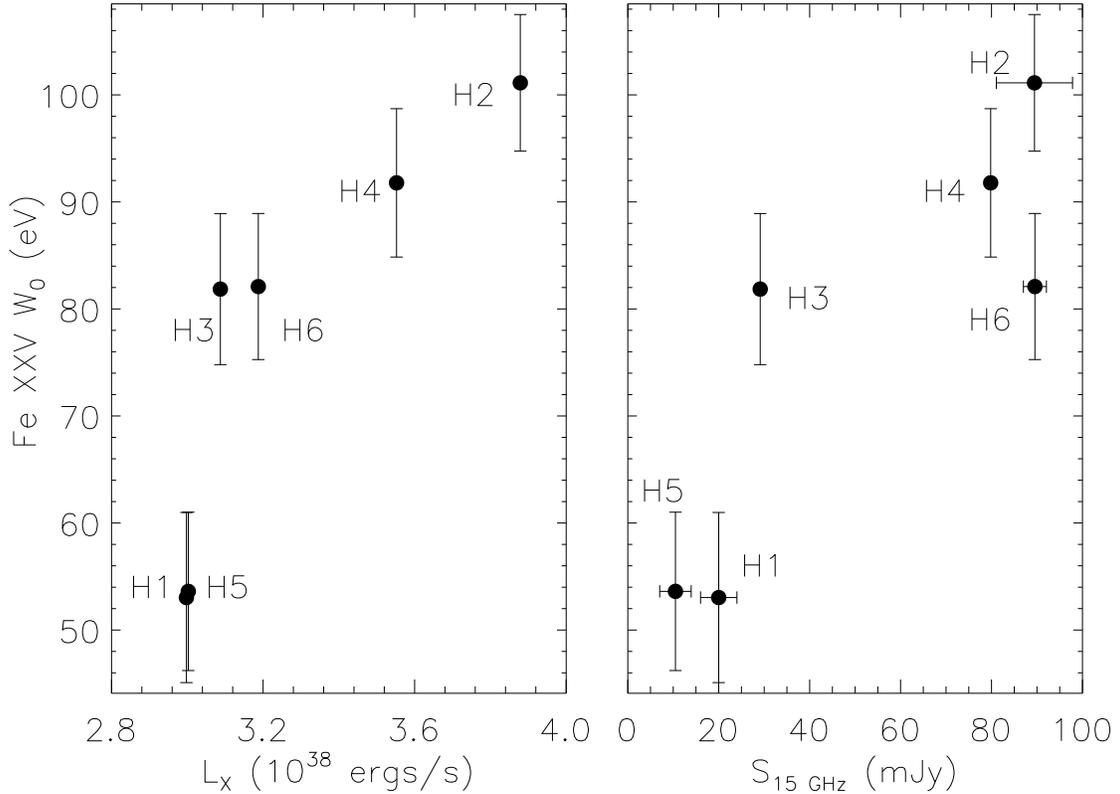}}
\vspace{-20mm}
\caption{The equivalent width $W_{0}$ of the broad Fe\,{\sc xxv}
  emission line in the hard states of GRS 1915+105 as a function of
  $L_{\rm X}$ (left) and radio flux (right). Because the emission line 
  equivalent width is correlated with both $L_{\rm X}$ and 
  $S_{\rm 15~GHz}$, it is more likely that the accretion disk is
  illuminated by the X-ray emitting base of the radio jet than by a
  hot, tenuous corona. Because the hard flux fraction does
  not obviously scale with $L_{\rm X},$ more detailed studies are
  required to determine the complex relationship between the X-ray
  illumination that causes the broad lines and the hard X-rays that
  ionize the wind. $W_{0}$ was measured based on simple Gaussian fits;
  the actual values will be larger if continuum uncertainties have
  masked the broad red wing of the line. Errors shown for $W_{0}$
  correspond to 90\% confidence limits on the emission line flux;
  errors shown for the radio flux at 15 GHz ($S_{\rm 15~GHz}$; ref 15)
  are also 90\% confidence limits.\\\\ } 
\label{fig3}
\end{figure}

\begin{figure}
\centerline{\includegraphics[width=\textwidth]{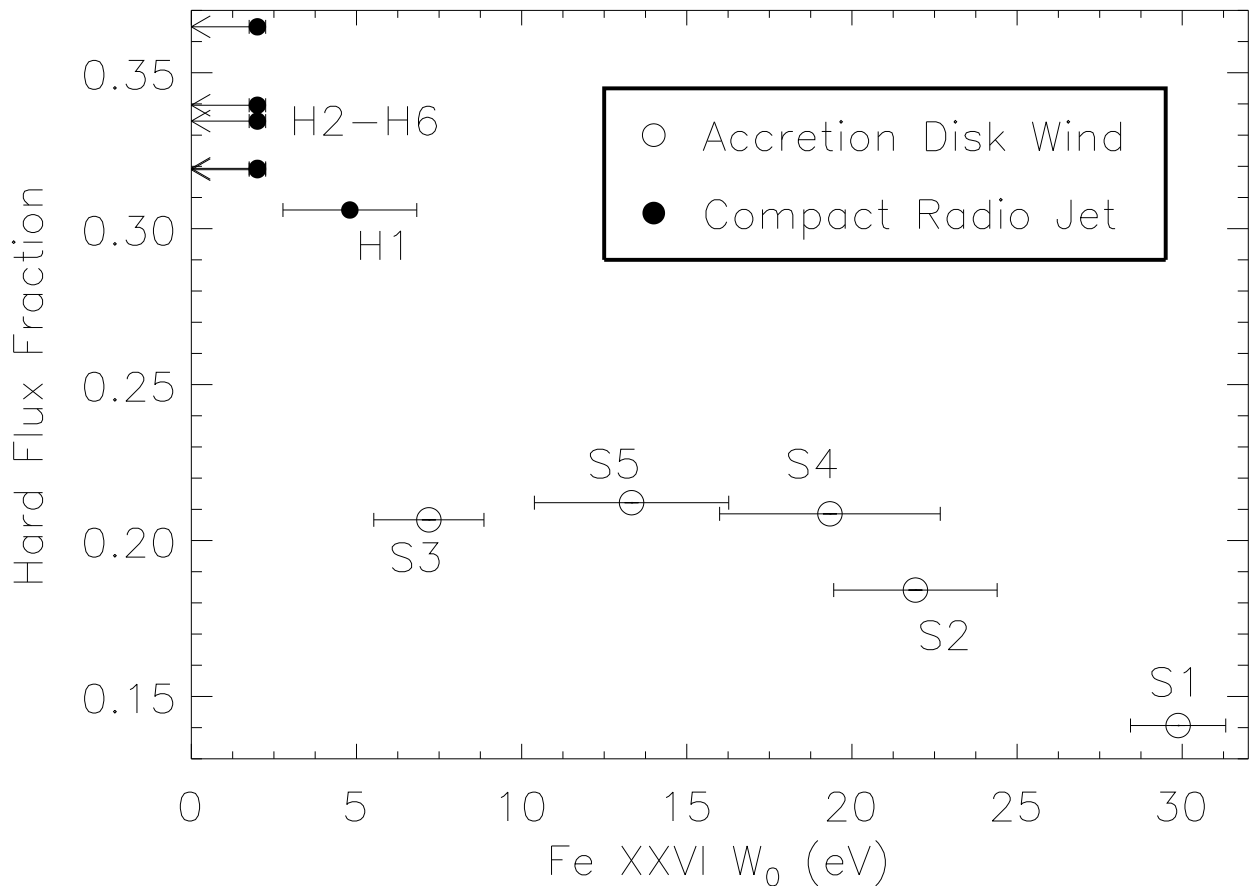}}
\vspace{-15mm}
\caption{The hard flux fraction and the equivalent width $W_{0}$ of
  the Fe\,{\sc xxvi} absorption line seen in GRS 1915+105. The
  hard flux fraction is defined as the ratio of the unabsorbed
  0.7--1.4 \AA~(8.6--18 keV) to 0.7--4.1 \AA~(3--18 keV) continuum
  fluxes, measured with RXTE. This figure shows $HF$ decreasing
  with $W_{0},$ implying that the jet weakens as the wind strength
  increases (and vice-versa). It appears that by carrying a
  significant amount of matter away from the accretion disk, strong
  winds can suppress jet production. This figure illustrates the
  nature of the complex competition between the wind and the jet,
  because the trend can also be understood in terms of the hard X-ray
  illumination of the wind. In the hardest states, the corona/jet may
  completely photoionize the wind, so that the gas is transparent, and
  therefore the absorption lines are weak or absent and the mass loss
  rate in the wind decreases significantly. This effect can also
  explain (for higher hard flux fraction) the disappearance of the
  lower-ionization absorption lines present in Obs.\ S1. Because of
  the complicated coupling between the disk and jet, we do not rule
  out broadened emission lines in softer states, or narrow absorption
  lines in harder states. In fact, the presence of both a broad
  emission \textit{and} narrow absorption lines in Obs.\ H1 provides
  further evidence of active competition between the wind and the
  jet. These results imply that $HF$ is a viable indicator of
  accretion physics around black holes. The errors shown for the
  equivalent width correspond to 90\% confidence limits on the
  absorbed line flux.}
\label{fig4}
\end{figure}
\clearpage
\begin{center}
\huge Supplementary Figure 1
\end{center}
\begin{figure}
\centerline{\includegraphics[width=\textwidth]{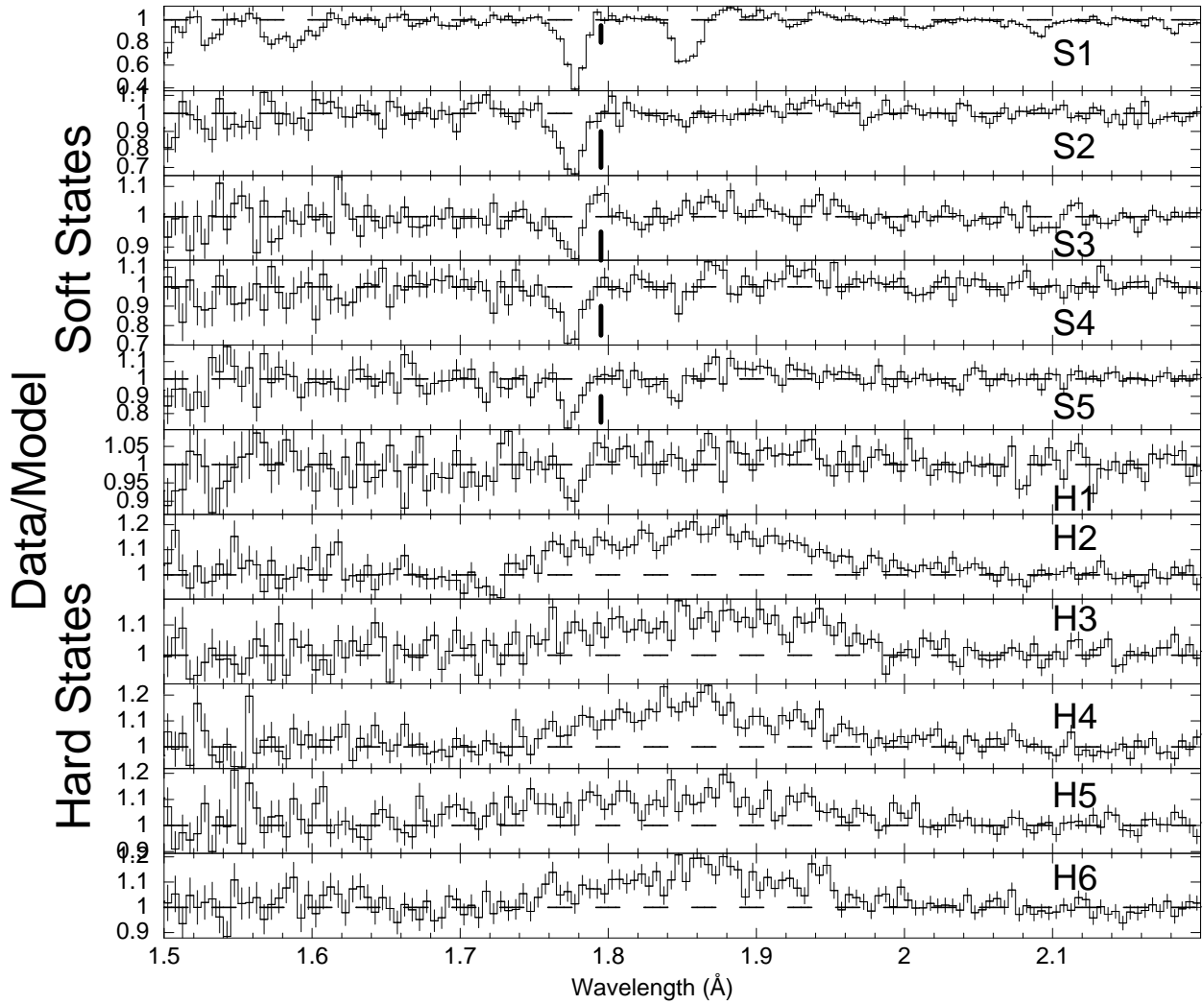}}
\label{fig1}
\vspace{-10mm}
\caption{Data:model ratio for all 11 HETGS observations of the
  microquasar GRS 1915+105. The spectra are plotted in order of
  increasing hard flux fraction (the ratio of the unabsorbed
  luminosity in the 8.6--18 keV band and the 3--18 keV band). `S'
  indicates a soft state and `H' indicates a hard state. The error
  bars shown are 1$\sigma$ statistical uncertainties on the data:model
  ratio. The vertical lines indicate the location of the emission
  component of the P-Cygni profiles.}  
\end{figure}


\begin{thebibliography}{1}
\bibitem{d1} Mirabel, I.~F. \& Rodriguez, L.~F. Sources of
  relativistic jets in the
  Galaxy. \textit{Annu. Rev. Astron. Astrophys.} \textbf{37}, 409-443
  (1999). 
\bibitem{d2} Fender, R.~P. \& Belloni, T.~GRS 1915+105 and the
  disc-jet coupling in accreting black hole
  systems. \textit{Annu. Rev. Astron. Astrophys.} \textbf{42}, 317-364
  (2004). 
\bibitem{d3} Klein-Wolt, M., et al. Hard X-ray states and radio emission in
  GRS 1915+105. \textit{Mon. Not. R. Astron. Soc.} \textbf{331},
  745-764 (2002). 
\bibitem{d4} Fender, R.,  et al. Quenching of the radio jet during the
  X-ray high state of GX 339-4. \textit{Astrophys. J.} \textbf{519},
  L165-L168 (1999). 
\bibitem{d5} Begelman, M.~C., McKee, C.~F., \& Shields, G.~A. Compton
  heated winds and coronae above accretion disks. I
  dynamics. \textit{Astrophys. J.} \textbf{271}, 70-88 (1983). 
\bibitem{d6} Proga, D. \& Kallman, T.~R. On the role of the
  ultraviolet and X-ray radiation in driving a disk wind in X-ray
  binaries. \textit{Astrophys. J.} \textbf{565}, 455-470 (2002).  
\bibitem{d7} Proga, D. Winds from accretion disks driven by radiation
  and magnetocentrifugal force. \textit{Astrophys. J.} \textbf{538},
  684-690 (2000).  
\bibitem{d8} Greiner, J., Cuby, J.~G., \& McCaughrean, M.~J. An
  unusually massive stellar black hole in the Galaxy. \textit{Nature},
  \textbf{414}, 522-524 (2001). 
\bibitem{d9} Mirabel, I.~F. \& Rodriguez, L.~F. A superluminal source
  in the Galaxy. \textit{Nature}, \textbf{371}, 46-48 (1994). 
\bibitem{d10} Belloni, T., Klein-Wolt, M., Mendez, M., van der Klis,
  M., \& van Paradijs, J. A model-independent analysis of the
  variability of GRS 1915+105. \textit{Astron. Astrophys.}
  \textbf{355}, 271-290 (2000). 
\bibitem{d11} Hannikainen, D. et al. Characterizing a new class of
  variability in GRS 1915+105 with simultaneous INTEGRAL/RXTE
  observations. \textit{Astron. Astrophys.} \textbf{435}, 995-1004
  (2005). 
\bibitem{d12} Mirabel, I.~F. et al. Accretion instabilities and jet
  formation in GRS 1915+105. \textit{Astron. Astrophys.} \textbf{330},
  L9-L12 (1998). 
\bibitem{d13} Eikenberry, S.~S., Matthews, K., Morgan, E.~H.,
  Remillard, R., \& Nelson, R.~W. Evidence for a disk-jet interaction
  in the microquasar GRS 1915+105. \textit{Astrophys. J.}
  \textbf{494}, L61-L64 (1998). 
\bibitem{d14} Fender, R.~P., Pooley, G.~G., Brocksopp, C., \& Newell,
  S.~J. Rapid infrared flares in GRS 1915+105: evidence for infrared
  synchrotron emission. \textit{Mon. Not. R. Astron. Soc.}
  \textbf{290}, L65-L69 (1997). 
\bibitem{d15} Pooley, G.~G., \& Fender, R.~P. The variable radio
  emission from GRS 1915+105. \textit{Mon. Not. R. Astron. Soc.}
  \textbf{292}, 925-933 (1997). 
\bibitem{d16} Canizares, C., et al. The Chandra high-energy transmission
  grating: design, fabrication, ground calibration, and 5 years in
  flight. \textit{Pub. Astron. Soc. Pacif.} \textbf{117}, 1144-1171
  (2005). 
\bibitem{d17} Dhawan, V., Mirabel, I.~F., \& Rodriguez, L.~F. AU-scale
  synchrotron jets and superluminal ejecta in GRS
  1915+105. \textit{Astrophys. J.} \textbf{543}, 373-385 (2000). 
\bibitem{d18} Kallman, T.~R. \& Bautista, M. Photoionization and
  high-density gas. \textit{Astrophys. J. Supp.} \textbf{133}, 221-253
  (2001). 
\bibitem{d19} Lee, J.C., et al. High-resolution Chandra HETGS and Rossi X-Ray
  Timing Explorer observations of GRS 1915+105: a hot disk atmosphere
  and cold gas enriched in iron and silicon. \textit{Astrophys. J.}
  \textbf{567}, 1102-1111 (2002). 
\bibitem{d20} Muno, M., Morgan, E.~H., \& Remillard, R. Quasi-periodic
  oscillations and spectral states in GRS
  1915+105. \textit{Astrophys. J.} \textbf{527}, 321-340 (1999).  
\bibitem{d21} Kotani, T., et al. ASCA observations of the absorption line
  features from the superluminal jet source GRS
  1915+105. \textit{Astrophys. J.} \textbf{539}, 413-423 (2000). 
\bibitem{d22} Esin, A.~A., McClintock, J.~E., \& Narayan,
  R. Advection-dominated accretion and the spectral states of black
  hole X-ray binaries: application to Nova Muscae
  1991. \textit{Astrophys. J.} \textbf{489}, 865-889 (1997). 
\bibitem{d23} Markoff, S., Nowak, M.~A., \& Wilms, J. Going with the
  flow: can the base of jets subsume the role of compact accretion
  disk coronae? \textit{Astrophys. J.} \textbf{635}, 1203-1216 (2005).  
\bibitem{d24} Eikenberry, S., et al. Spectroscopy of infrared flares from
  the microquasar GRS 1915+105. \textit{Astrophys. J.} \textbf{506},
  L31-L34 (1998). 
\bibitem{d25} McClintock, J., et al. The spin of the near-extreme Kerr
  black hole GRS 1915+105. \textit{Astrophys. J.} \textbf{652},
  518-539 (2006). 
\bibitem{d26} Proga, D., Stone, J.~M., \& Kallman, T.~R. Dynamics of
  line-driven winds in active galactic nuclei. \textit{Astrophys. J.}
  \textbf{543}, 686-696 (2000). 
\bibitem{d27} Janiuk, A., Czerny, B., \& Siemiginowska, A. Radiation
  pressure instability as a variability mechanism in the microquasar
  GRS 1915+105. \textit{Astrophys. J.} \textbf{542}, L33-L36 (2000). 
\bibitem{d28} Houck, J.~C. \& Denicola, L.~A. in \textit{Astronomical
    Data Analysis Software and Systems IX} (eds Manset, N., Veillet,
  C. \& Crabtree, D.) 591-594 (ASP Conference Series Vol. 216, 2000).
\end{thebibliography}
\end{document}